\documentclass
[notitlepage,12pt,groupedaddress,secnumarabic,prd,showpacs,showkeys,onecolumn]{revtex4}%
\usepackage{amssymb}
\usepackage{amsfonts}
\usepackage{amsmath}
\usepackage{latexsym}
\usepackage{graphicx}
\usepackage{appendix}%
\begin{document}
\title{Light wave propagation through a dilaton-Maxwell domain wall\\ }
\author{J.R. Morris$^{1,a)}$ and A. Schulze-Halberg$^{2,b)}$}
\affiliation{$^{1}$Department of Physics, Indiana University Northwest, 3400 Broadway,
Gary, Indiana, 46408, USA, $^{2}$Department of Mathematics and Department of
Physics, Indiana University Northwest, 3400 Broadway, Gary, Indiana, 46408, USA}
\email{a) jmorris@iun.edu,b) axgeschu@iun.edu, xbataxel@gmail.com}

\begin{abstract}
We consider the propagation of electromagnetic waves through a dilaton-Maxwell
domain wall of the type introduced by Gibbons and Wells [G.W. Gibbons and C.G.
Wells, Class. Quant. Grav. 11, 2499-2506 (1994)]. It is found that if such a
wall exists within our observable universe, it would be absurdly thick, or
else have a magnetic field in its core which is much stronger than observed
intergalactic fields. We conclude that it is highly improbable that any such
wall is physically realized.

\end{abstract}

\maketitle

\section{Introduction}

\ \ Starting from a five dimensional Kaluza-Klein theory, which is toroidally
compactified to yield an effective four dimensional dilaton-Maxwell theory, we
find exact background solutions describing a dilatonic domain wall which
entraps magnetic flux, which has previously been described by Gibbons and
Wells \cite{GW}. This type of domain wall is interesting, not only because it
traps magnetic flux, but also because it is nontopological in origin, i.e.,
the solution is not stabilized by a nontrivial topology of the vacuum
manifold. (However, this stability issue was examined in \cite{wall}, where it
was determined that the Gibbons-Wells wall is indeed stable against small
fluctuations in the scalar and magnetic fields.) Particles, including both
fermions and bosons, can scatter from a topological domain wall in various
ways (see \cite{Vilenkin}-\cite{Rajaraman} for example). In particular, the
scattering of scalar bosons from such walls has been examined in
\cite{section}. In addition, a coupling of a scalar dilaton field (with a
simple quartic potential) to matter and electromagnetic fields has been
studied in \cite{Olive11}, where it was proposed that the existence of a
dilaton domain wall might give rise to spatial variations in the fine
structure constant $\alpha$. We do not concern ourselves with specifics of
such a type of scenario involving a Gibbons-Wells wall, but simply point out
that interactions of dilatonic walls with matter and electromagnetic fields
may indeed be of physical importance.

\bigskip

\ \ The main focus here is on the propagation of electromagnetic waves in the
dilaton-Maxwell domain wall. Exact solutions for the wave equation are found,
and it is determined that there is a critical frequency above which there are
transmitted travelling waves which are damped in amplitude as the distance
$|x|$ from the core of the wall tends to infinity. We argue, however, that the
wall is transparent to essentially all electromagnetic waves if the effective
dielectric function is to have very small spatial variation. We speculate on
some observable consequences of the existence of such a domain wall. The
existence of such a solitonic structure would support the possibility of the
existence of extra compactified space dimensions. However, what we infer is
that the wall's magnetic field is either too large in comparison to an
intergalactic field strength, or else the thickness of the wall is absurdly
large. We conclude that it seems improbable that a Gibbons-Wells wall is
physically realized in our observable universe.

\section{The dilaton-Maxwell domain wall}

\subsection{Equations of motion}

\ \ We start with a 5d action, using a 5d metric described by $d\tilde{s}%
_{5}^{2}=\tilde{g}_{\mu\nu}dx^{\mu}dx^{\nu}-b^{2}(x^{\mu})dy^{2}$ with
signature $(+,-,-,-,-)$ which is dimensionally reduced by toroidal
compactification and rewritten in a 4d Einstein conformal frame, subsequently
taking the form (see, e.g., Ref. \cite{JM03} for details)%
\begin{equation}
S=\int d^{4}x\sqrt{g}\Big\{\frac{1}{2\kappa^{2}}R[g_{\mu\nu}]+\frac{1}%
{2}\partial^{\mu}\varphi\partial_{\mu}\varphi+e^{2\tilde{\kappa}\varphi
}\left(  \mathcal{L}-\frac{\Lambda}{\kappa^{2}}\right)  \Big\} \label{1}%
\end{equation}

where $\kappa^{2}=8\pi G$, $g=|\det g_{\mu\nu}|$ with $g_{\mu\nu}$ being the
Einstein metric, and $\mathcal{L}$ is the matter lagrangian written in terms
of the 4d Jordan frame metric $\tilde{g}^{\mu\nu}$. Here $2\tilde{\kappa
}=-\sqrt{2/3}\kappa$ is a constant, and we have assumed $\partial_{5}%
=\partial/\partial y=0$ (no Kaluza-Klein modes), and we use natural units with
$\hbar=c=k_{B}=1$. The extra dimensional scale factor is denoted by
$b=e^{-2\tilde{\kappa}\varphi}$ and the relation between the 4d Jordan frame
metric $\tilde{g}_{\mu\nu}$ and the 4d Einstein frame metric $g_{\mu\nu}$ is
given by%
\begin{equation}
g_{\mu\nu}=b\tilde{g}_{\mu\nu}=e^{-2\tilde{\kappa}\varphi}\tilde{g}_{\mu\nu
},\ \tilde{g}^{\mu\nu}=e^{-2\tilde{\kappa}\varphi}g^{\mu\nu} \label{2}%
\end{equation}

The lagrangian forming $\mathcal{L}$ is taken to be
\begin{equation}
\mathcal{\mathcal{L}}=-\frac{1}{4}\tilde{F}^{\mu\nu}\tilde{F}_{\mu\nu}%
=-\frac{1}{4}e^{-4\tilde{\kappa}\varphi}F^{\mu\nu}F_{\mu\nu}=-\frac{1}%
{2}e^{-4\tilde{\kappa}\varphi}(\mathbf{B}^{2}-\mathbf{E}^{2}) \label{3}%
\end{equation}

where $\tilde{F}^{\mu\nu}=\tilde{g}^{\mu\alpha}\tilde{g}^{\nu\beta}%
F_{\alpha\beta}$, with $\tilde{F}_{\mu\nu}=F_{\mu\nu}=\partial_{\mu}A_{\nu
}-\partial_{\nu}A_{\mu}$. Also, following \cite{GW}, we focus on the flat
space version where the Ricci scalar $R[g_{\mu\nu}]=0$ and the metric is
Minkowski, $g_{\mu\nu}=\eta_{\mu\nu}$, and we set the cosmological constant to
zero, $\Lambda=0$.

\bigskip

\ \ Now, the equations of motion that follow from (\ref{1}), along with the
Bianchi identity, are given by%
\begin{equation}
\square\varphi+\frac{1}{2}\tilde{\kappa}e^{-2\tilde{\kappa}\varphi}F^{\mu\nu
}F_{\mu\nu}=0 \label{5}%
\end{equation}

\begin{equation}
\nabla_{\mu}(e^{-2\tilde{\kappa}\varphi}F^{\mu\nu})=0,\ \ \ \nabla_{\mu
}\ (^{\ast}F^{\mu\nu})=0 \label{6}%
\end{equation}

where $\square=\partial_{t}^{2}-\nabla^{2}$ and the electromagnetic dual
tensor is $^{\ast}F_{\mu\nu}=\frac{1}{2}\epsilon_{\mu\nu\rho\sigma}%
F^{\rho\sigma}$. The set of equations in (\ref{6}) is just the set of Maxwell
equations%
\begin{equation}
\nabla\cdot\mathbf{D}=0,\ \ \nabla\times\mathbf{H}-\mathbf{\dot{D}%
}=0,\ \ \nabla\cdot\mathbf{B}=0,\ \ \nabla\times\mathbf{E}+\mathbf{\dot{B}}=0
\label{8}%
\end{equation}

with $\mathbf{D}=\epsilon\mathbf{E}$ and $\mathbf{B}=\mu\mathbf{H}$, where the
effective dielectric and permeability functions are $\epsilon=\mu^{-1}$ with%
\begin{equation}
\mu=\epsilon^{-1}=e^{2\tilde{\kappa}\varphi} \label{9}%
\end{equation}

and the index of refraction is $\sqrt{\epsilon\mu}=1$. We can rewrite
(\ref{5}) now in terms of $\mathbf{D}$ and $\mathbf{H}$:%
\begin{equation}
\nabla^{2}\varphi-\partial_{t}^{2}\varphi=-\tilde{\kappa}e^{2\tilde{\kappa
}\varphi}(\mathbf{H}^{2}-\mathbf{D}^{2}) \label{10}%
\end{equation}

\subsection{The background ansatz}

\ \ An exact, static solution set can be found for the above equations of
motion. This solution set then serves as a background for the scattering of
electromagnetic (EM) waves. The background solution is that of a dilatonic
domain wall entrapping magnetic flux, originally discovered by Gibbons and
Wells \cite{GW}. For this static solution, we set $\mathbf{D}=0$,
$\mathbf{E}=0$, $\mathbf{H}=(0,0,H)$, where $H$ is a constant, and
$\mathbf{B}=(0,0,B)=\mu\mathbf{H}$. We find that the Maxwell equations
(\ref{8}) are then satisfied, and the equation of motion for the dilaton field
$\varphi$ of (\ref{10}) then reduces to%
\begin{equation}
(\partial_{x}^{2}+\partial_{y}^{2})\varphi=-\tilde{\kappa}H^{2}e^{2\tilde
{\kappa}\varphi} \label{11}%
\end{equation}

where we assume that $\partial_{z}\varphi=0$. This equation is recognized as
the 2D Euclidean Liouville equation \cite{Liouville} whose solution is given
by \cite{GW,Liouville,Crowdy,wall}%
\begin{equation}
\mu(\zeta)=e^{2\tilde{\kappa}\varphi(\zeta)}=\frac{4}{\tilde{\kappa}^{2}H^{2}%
}\frac{\left\vert f^{\prime}(\zeta)\right\vert ^{2}}{\left(  1+\left\vert
f(\zeta)\right\vert ^{2}\right)  ^{2}} \label{12}%
\end{equation}

where $\zeta=x+iy$ and $f(\zeta)$ is a holomorphic function of $\zeta$ and
$f^{\prime}(\zeta)=df(\zeta)/d\zeta$. Let us choose $f(\zeta)$ to take the
form $f(\zeta)=e^{M\zeta}$. Then (\ref{12}) produces a static domain wall
solution \cite{GW,JM06,wall}%
\begin{equation}
\mu(x)=e^{2\tilde{\kappa}\varphi(x)}=\left(  \frac{M}{\tilde{\kappa}H}\right)
^{2}\frac{1}{\cosh^{2}(Mx)}=\left(  \frac{M}{\tilde{\kappa}H}\right)
^{2}\text{sech}^{2}(\bar{x}),\ \ \ \ \bar{x}\equiv Mx \label{13}%
\end{equation}

The constant $M$ has a canonical dimension of mass, so that the coordinate
$\bar{x}=Mx$ is dimensionless, as is the factor $(M/\tilde{\kappa}H)$. This
domain wall solution depends only upon $x$, and not upon $y$, and the width of
the wall is represented by $a=M^{-1}$. The magnetic field is $B(x)=\mu
(x)H\propto$ sech$^{2}\bar{x}$, which maximizes in the wall's core and falls
to zero asymptotically. The magnetic flux per unit length of the domain wall
is \cite{GW,JM06,wall}
\begin{equation}
\frac{\Phi_{\text{mag}}}{L_{y}}=\frac{1}{L_{y}}\int_{-\infty}^{\infty}\int
_{0}^{L_{y}}B(x)dxdy=\frac{2M}{\tilde{\kappa}^{2}H} \label{14}%
\end{equation}

\section{Electromagnetic wave propagation}

\subsection{Wave equations and exact solutions}

\ \ We now examine the scattering of electromagnetic (EM) waves from the wall
background ansatz of (\ref{13}), except now we denote the static magnetic $B$
and $H$ fields of the wall by $B_{0}$ and $H_{0}$, and denote those of
electromagnetic waves by $B$ and $H$. \ The basic formalism for EM scattering
from a dilatonic wall (with normal incidence) with arbitrary $\epsilon(x)$ and
$\gamma(x)=\ln\epsilon(x)$ is described in Sec.IVa of \cite{RT}. We use
results presented there to describe EM wave fields with nonvanishing
components $E_{y}(x,t)$ and $B_{z}(x,t)$ propagating in the $\pm x$ direction.
The electromagnetic field equations can be reduced to \cite{RT}%
\begin{equation}
\partial_{x}^{2}B_{z}-\partial_{t}^{2}B_{z}+(\partial_{x}^{2}\gamma
)B_{z}+(\partial_{x}\gamma)\partial_{x}B_{z}=0 \label{e1}%
\end{equation}

and we assume fields of the form%
\begin{equation}
E_{y}(x,t)=E(x,\omega)e^{-i\omega t},\ \ \ \ B_{z}(x,t)=B(x,\omega)e^{-i\omega
t} \label{e2}%
\end{equation}

From the field equations we then have%
\begin{equation}
E=-\frac{i}{\omega}\left[  \partial_{x}B+(\partial_{x}\gamma)B\right]
\label{e3}%
\end{equation}
We again define the dimensionless coordinate $\bar{x}=Mx$, and using
$\partial_{x}\gamma=2M\tanh\bar{x}$ and $\partial_{x}^{2}\gamma=2M^{2}%
$sech$^{2}\bar{x}$, (\ref{e1}) can be written as%
\begin{equation}
\partial_{\bar{x}}^{2}B+2(\tanh\bar{x})\partial_{\bar{x}}B+\left(
\frac{\omega^{2}}{M^{2}}+2\text{sech}^{2}\bar{x}\right)  B=0 \label{e4}%
\end{equation}

Now we change coordinates according to
\begin{equation}
\bar{x}(u)~=\text{arctanh}(u),\qquad B[\bar{x}(u)]~=~\sqrt{u^{2}-1}~\hat
{B}(u). \label{a2}%
\end{equation}
This renders our equation (\ref{e4}) in the form
\begin{equation}
(1-u^{2})^{2}~\hat{B}^{\prime\prime}(u)-2~u~\hat{B}^{\prime}(u)+\left[
\frac{M^{2}~(2~u^{2}-1)-\omega^{2}}{M^{2}~(u^{2}-1)}\right]  \hat{B}(u)=0.
\label{a3}%
\end{equation}
where here the prime denotes differentiation with respect to the argument $u$.
The general solution to this equation is given by
\begin{equation}
\hat{B}(u)=c_{1}~P_{1}^{\mu}(u)+c_{2}~Q_{1}^{\mu}(u),\qquad\mu~=~\sqrt
{1-\frac{\omega^{2}}{M^{2}}} \label{a4}%
\end{equation}
After reverting the change of coordinate (\ref{a2}), the latter solution takes
the form
\begin{equation}
B=\text{sech}(\bar{x})\left[  c_{1}(\omega)P_{1}^{\mu}(\xi)+c_{2}(\omega
)Q_{1}^{\mu}(\xi)\right]  \label{e5}%
\end{equation}

where $P$ and $Q$ are Legendre functions, $c_{1}(\omega)$ and $c_{2}(\omega)$
are $\bar{x}$ independent parameters, which, in general, can depend upon the
frequency $\omega$, and%
\begin{equation}
\mu=\sqrt{1-\frac{\omega^{2}}{M^{2}}},\ \ \ \ \ \xi=\tanh\bar{x} \label{e6}%
\end{equation}

where the index $\mu$ of (\ref{e6}) is not to be confused with the
permeability function defined earlier. Due to the lower index on $P$ and $Q$
being integer, the solution (\ref{e5}) degenerates to an elementary function,
which is given by%
\begin{align}
B  &  =\text{sech}(\bar{x})\Big\{c_{1}\frac{(\xi-\mu)(1+\xi)^{\mu/2}}%
{\Gamma(2-\mu)(1-\xi)^{\mu/2}}\nonumber\\
&  +c_{2}\frac{\pi(1-\xi^{2})^{-\mu/2}\left[  (1+\xi)^{\mu}(\xi-\mu)\cos
(\pi\mu)-(1-\xi)^{\mu}(\xi+\mu)\right]  }{2\sin(\pi\mu)\Gamma(2-\mu
)}\Big\} \label{e7}%
\end{align}

The behavior of the solution (\ref{e7}) can be described as follows. First we
observe that the factor sech$(\bar{x})$ is responsible for damping, as
$|\bar{x}|$ tends toward infinity. The term inside the curly brackets is in
general complex-valued. Its qualitative behavior depends principally on the
quantity $\mu$, which depends upon $\omega/M$. We can distinguish two cases,
assuming without restriction that $c_{1}$ and $c_{2}$ are real-valued.

\begin{description}
\item (1)\ \ $\boldsymbol{\omega\leq M},$ \textbf{i.e.,} \textbf{real root,}
$\boldsymbol{(1-\omega}^{2}\boldsymbol{/M}^{2}\boldsymbol{)\geq0:}$ The
solution (\ref{e7}) is real, bounded everywhere, has two zeros (one positive,
one negative) and goes to zero as $|x|$ tends to infinity. These properties
are independent of $c_{1}$ and $c_{2}$. These solutions are non-oscillatory
outside of the domain wall.

\item (2)\ \ $\boldsymbol{\omega>M,}$ \textbf{i.e.,} \textbf{imaginary root,}
$\boldsymbol{1-\omega}^{2}\boldsymbol{/M}^{2}\boldsymbol{<0:}$ The solution
(\ref{e7}) is complex. Both real and imaginary parts are bounded everywhere,
have an infinite number of zeros and go to zero as $|\bar{x}|$ tends to
infinity. The set of zeros is unbounded from both below and above. These
properties are independent of $c_{1}$ and $c_{2}$.
\end{description}

\subsection{An estimate for the domain wall mass parameter $M$ and width
$a=M^{-1}$}

\ \ Since dilatonic-matter effects are expected to be nearly negligible and
nearly undetectable at this time, we focus on the case where the effective
dielectricity in vacuum $\epsilon(\bar{x})$ does not wander far from unity. We
define $\epsilon_{0}=1$ for the case of no dilaton coupling to EM fields,
i.e., ordinary electrodynamics, and consider the case where $\Delta
\epsilon(\bar{x})=\epsilon(\bar{x})-\epsilon_{0}\ll1$. Furthermore, we
consider the possibility where there is just one dilaton-Maxwell wall within
the observable universe, and we roughly estimate the width of the wall to be
on the order of the Hubble length, $a\sim|x_{C}|\sim l_{H}\sim10^{10}$ light
years $\sim10^{26}$ m. The distance $|x_{C}|$ serves as a long distance
cutoff, as the wall's surface energy density (tension) $\Sigma(x)$ diverges
with distance away from the wall according to \cite{wall} $\Sigma
(|x|)=(M/\tilde{\kappa}^{2})|x|$, so that%
\begin{equation}
\Sigma(|x_{C}|)=\frac{M}{\tilde{\kappa}^{2}}|x_{C}| \label{e8}%
\end{equation}

\ \ Our expression for\ the effective dielectric constant is%
\begin{equation}
\epsilon(\bar{x})=e^{-2\tilde{\kappa}\varphi}=\left(  \frac{\tilde{\kappa
}H_{0}}{M}\right)  ^{2}\cosh^{2}(\bar{x}) \label{e9}%
\end{equation}

where $-2\tilde{\kappa}=\sqrt{2/3}\kappa$ and $\bar{x}=Mx=x/a$. We therefore
examine the limit where $\kappa|\varphi|\ll1$ and $\epsilon(\bar{x}%
)\ll1+\epsilon_{0}=2$. We therefore want to consider $|\bar{x}_{C}|$ to be not
far from order unity so that the $\cosh^{2}(\bar{x})$ term has little
variance, even over large distances. (We expect standard classical EM theory
and QED to hold to a high degree of precision everywhere within the observable
universe, with dilatonic effects being extremely small.) We choose to set
$|\frac{\tilde{\kappa}H_{0}}{M}|=1$, which fixes $H_{0}$ in terms of $M$, so
that%
\begin{equation}
\epsilon(\bar{x})=\cosh^{2}(\bar{x})=1+\bar{x}^{2}+O(\bar{x}^{4})\approx
1+\bar{x}^{2} \label{e10}%
\end{equation}

We then have that $\Delta\epsilon(\bar{x})=\epsilon(\bar{x})-1=\bar{x}%
^{2}=(Mx)^{2}\ll1$. Setting $|\bar{x}|=|\bar{x}_{C}|=M|x_{C}|\lesssim1$ we
have $M\lesssim1/|x_{C}|$ and therefore a wall thickness $a=M^{-1}%
\gtrsim|x_{C}|\sim l_{H}\sim10^{26}$ m, where $l_{H}$ is the Hubble length. So
for $a\sim|x_{C}|$ on the order of the Hubble length, we have a very thick
wall, extending through the observable universe. EM radiation of essentially
all wavelengths are much smaller than the wall thickness, i.e., $\lambda\ll
a=M^{-1}\sim l_{H}$ for essentially all radiation with wavelength smaller than
the Hubble length, and therefore we have $\omega/M>1$ for essentially all
radiation. Therefore, all radiation travels through a very thick wall
($\lambda\ll a=M^{-1}$) where $\epsilon(x)$ varies very slowly.

\section{Wave propagation through a thick wall}

\subsection{Wave equations and approximate solutions}

\ \ The propagation of EM waves through a thick wall with arbitrary, but
slowly varying, $\epsilon(x)$ has been described in \cite{RT}. Instead of
repeating the calculational details presented there, we simply recap some of
the highlights. We define the function $\gamma(x)=\ln\epsilon(x)$ and consider
$\partial_{x}\epsilon$ and $\partial_{x}\gamma$ to be sufficiently small.\ Let
us now examine the scattering of electromagnetic (EM) waves from the wall
background ansatz of the dilaton-Maxwell wall, except now we denote the static
magnetic $B$ and $H$ fields of the wall by $B_{0}$ and $H_{0}$, and denote
those of electromagnetic waves by $B$ and $H$. \ The basic formalism for EM
scattering from a dilatonic wall (with normal incidence) with arbitrary
$\epsilon(x)$ and $\gamma(x)=\ln\epsilon(x)$ is described in Sec.IVa of
\cite{RT}, and the reader is referred there for calculational details. We use
results presented there to describe EM wave fields with nonvanishing
components $E_{y}(x,t)$ and $B_{z}(x,t)$ propagating in the $\pm x$ direction.
For notational simplicity, we simply write $E=E_{y}$ and $B=B_{z}$. The
electromagnetic field equations can be reduced to \cite{RT}%
\begin{equation}
B^{\prime\prime}-\ddot{B}+\gamma^{\prime\prime}B+\gamma^{\prime}B^{\prime
}=0,\ \ \ \ \ E=-\frac{i}{\omega}\left[  B^{\prime}+\gamma^{\prime}B\right]
\label{e11}%
\end{equation}

where we assume the fields to have the time dependence $e^{-i\omega t}$ and a
prime stands for differentiation with respect to $x$.

\bigskip

\ \ We now take the magnetic field $B(x,t)$ to be of the form%
\begin{equation}
B(x,t)=Ae^{i\phi(x)}e^{-i\omega t} \label{e12}%
\end{equation}

where the amplitude $A$ is a real constant and $\phi(x)$ is a phase function,
which may be complex-valued, in general. The wave equation for $B$ then gives
an equation for the function $\phi$,%
\begin{equation}
i\phi^{\prime\prime}-\phi^{\prime2}+\omega^{2}+\gamma^{\prime\prime}%
+i\phi^{\prime}\gamma^{\prime}=0 \label{e13}%
\end{equation}

For the expectedly small dilatonic effect on the wave equations for $E$ and
$B$, we note that for the case $\epsilon=$ const and $\gamma^{\prime}=0$ we
have the usual solution $\phi^{\prime}=\pm\omega$ and $\phi=\pm\omega x$. The
$+(-)$ solution describes waves traveling in the $+x(-x)$ direction.
Approximations can be made for the case of slowly varying $\epsilon(x)$ which
lead to the approximate solutions for the EM fields \cite{RT}%
\begin{equation}
B_{\pm}(x,t)=A\left(  \frac{\epsilon}{\epsilon_{0}}\right)  ^{-1/2}e^{\pm
i\omega x}e^{-i\omega t} \label{e14}%
\end{equation}

and%
\begin{equation}
E_{\pm}(x,t)=\left(  \pm1-i\frac{\gamma^{\prime}}{2\omega}\right)  A\left(
\frac{\epsilon}{\epsilon_{0}}\right)  ^{-1/2}e^{\pm i\omega x}e^{-i\omega t}
\label{e15}%
\end{equation}

where $\epsilon_{0}$ is just a constant which can be set to unity. Note that
the effective amplitude of the magnetic field is $A\left(  \frac{\epsilon
}{\epsilon_{0}}\right)  ^{-1/2}$. Since we assume that $\Delta\epsilon\ll1$,
the effective amplitude varies only mildly over any distance of interest.

\subsection{Reflection and transmission coefficients}

\ \ The Poynting vector is given by $\mathbf{S}=\operatorname{Re}%
(\mathbf{E}\times\mathbf{H}^{\ast})=\epsilon\operatorname{Re}(\mathbf{E}%
\times\mathbf{B}^{\ast})$. This can be applied to each of the $\pm$
propagating waves, which after some algebra, yields \cite{RT}%
\begin{equation}
(S_{x})_{\pm}=\epsilon\operatorname{Re}(\mathbf{E}_{\pm}\times\mathbf{B}_{\pm
}^{\ast})=\pm\epsilon|\mathbf{B}_{\pm}|^{2}=\pm\epsilon\left[  A^{2}%
\frac{\epsilon_{0}}{\epsilon}\right]  =\pm\epsilon_{0}A^{2} \label{e16}%
\end{equation}

This shows that $(S_{x})_{\pm}$ is $x$ independent, which by Poynting's
theorem, indicates that no energy or momentum is lost by either the $+$ or $-$
traveling waves, which in turn implies that the transmission and reflection
coefficients are given by $T=1$ and $R=0$, respectively, for waves with
$\omega\gg|\gamma^{\prime}|=M|\tanh(\bar{x})|$, i.e., $\omega/M\gg|\tanh
(\bar{x})|$. We have argued that, within the scenario considered here,
$\omega/M>1$ for all EM waves of interest, and in fact, due to the extreme
smallness of $M$, we have $\omega/M\gg1$ for all frequencies of interest. The
wall is therefore transparent to all EM radiation.

\section{Observational consequences}

\ \ Since $T=1$ and $R=0$, it would therefore seem that there would be few
observational consequences. However, we recall that the setting $|\frac
{\tilde{\kappa}H_{0}}{M}|=1$ will fix the value of the wall-entrapped $H_{0}$
field in terms of the mass parameter $M$. We therefore have%
\begin{equation}
|H_{0}|=\frac{M}{|\tilde{\kappa}|}=\frac{2M}{\sqrt{2/3}\kappa}=2\sqrt{\frac
{3}{2}}\frac{M}{\kappa}=2\sqrt{\frac{3}{2}}\frac{M}{\sqrt{8\pi G}}\sim MM_{P}
\label{e17}%
\end{equation}

where $M_{P}=1/\kappa=1/\sqrt{8\pi G}\sim10^{18}$ GeV is the reduced Planck
mass. Our estimate of $M^{-1}\sim10^{26}$ m gives $M\sim10^{-26}$ m$^{-1}%
\sim10^{-42}$ GeV, so that%
\begin{equation}
|H_{0}|\sim MM_{P}\sim10^{-24}\ \text{GeV}^{2}\sim10^{-4}\text{\ G}
\label{e18}%
\end{equation}

where we have used the conversion that one Tesla (T) is $10^{4}$ Gauss(G),
approximately given by $1$ T $=10,000\ $G$\ \sim200$ eV$^{2}$ , or 1
GeV$^{2}\sim10^{16}$ T $\sim10^{20}$ G. This value of $|H_{0}|$ in the center
of the wall is much larger than an intergalactic magnetic field strength of
$\sim10^{-15}$ G \cite{Essey}. Therefore, setting $a\sim l_{H}$ results in a
magnetic field in space that is larger than that observed in intergalactic regions.

\bigskip

\ \ If, on the other hand, we set $|H_{0}|\sim10^{-15}$ G $\sim10^{-35}$
GeV$^{2}$ and use $M\sim|H_{0}|/M_{P}$, we can determine $a=M^{-1}$. Doing
this, we have%
\begin{equation}
M\sim\frac{|H_{0}|}{M_{P}}\sim10^{-53}\text{ GeV},\ \ \ a=M^{-1}\sim
10^{53}\text{ GeV}^{-1}\sim10^{37}\text{ m} \label{e19}%
\end{equation}

Since a characteristic Hubble length is $l_{H}\sim10^{26}$ m, this estimate
gives $a\sim10^{11}l_{H}$, i.e., $10^{11}$ Hubble lengths! In this case the
wall is ridiculously thick, spanning many observable universes.

\section{Summary and conclusions}

\ \ In summary, it is found that for a single dilaton-Maxwell wall, the
setting of $|\frac{\tilde{\kappa}H_{0}}{M}|=1$ along with $a\sim l_{H}$
results in a magnetic field that is too strong, while setting $|H_{0}|$ to the
value of an intergalactic magnetic field results in a domain wall spanning
many observable universes, with $a\sim10^{11}l_{H}$. On the other hand, if
$|\frac{\tilde{\kappa}H_{0}}{M}|\neq1$, we have a value of $\epsilon(x)$ that
wanders too far from unity in at least some regions of space, even if
$\cosh^{2}(\bar{x})\sim1$, which does not seem to be supported by observation.
If there were many thinner walls where $\epsilon(x)$ could differ from unity
for $|x|/a>1$, one would expect some observed periodic spatial variation in
the fine structure constant $\alpha$, which does not seem reasonable.
(However, other topological dilaton domain wall models that do not require a
large distance cut off may possibly allow for a mild variation of $\alpha$
\cite{Olive11}.) We conclude that a dilaton-Maxwell domain wall is not likely
to be physically realized within our observable universe.

\end{document}